\documentclass[12pt,a4paper,final]{iopart}
\usepackage{graphicx}
\usepackage{dsfont}
\usepackage{epstopdf}
\usepackage[breaklinks=true,colorlinks=true,linkcolor=blue,urlcolor=blue,citecolor=blue]{hyperref}
\usepackage{color}
\usepackage{float}
\usepackage{iopams}
\usepackage{array}

\newcolumntype{L}{>{$}l<{$}}
\graphicspath{{figs/}}
\begin{document}

\title[Probing quasi-integrability of the GPE in a harmonic-oscillator potential]{%
Probing quasi-integrability of the Gross-Pitaevskii equation in a
harmonic-oscillator potential}
\author{T. Bland$^1$, N. G. Parker$^1$, N. P. Proukakis$^1$ and \\ B. A. Malomed$^{2,3}$}

\address{$^1$Joint Quantum Centre Durham--Newcastle, School of Mathematics,
Statistics and Physics, Newcastle University, Newcastle upon Tyne, NE1 7RU,
United Kingdom}
\ead{\mailto{thomas.bland@ncl.ac.uk, nick.parker@ncl.ac.uk,
nick.proukakis@ncl.ac.uk}}

\address{$^2$Department of Physical Electronics, School of Electrical
Engineering, Faculty of Engineering, Tel Aviv University, Tel Aviv 69978,
Israel\\ $^3$ITMO University, St. Petersburg 197101, Russia} %
\ead{malomed@post.tau.ac.il}

\begin{abstract}
Previous simulations of the one-dimensional Gross-Pitaevskii equation (GPE)\
with repulsive nonlinearity and a harmonic-oscillator trapping potential
hint towards the emergence of quasi-integrable dynamics -- in the sense of
quasi-periodic evolution of a moving dark soliton without any signs of
ergodicity -- although this model does not belong to the list of integrable
equations. To investigate this problem, we replace the full GPE by a
suitably truncated expansion over harmonic-oscillator eigenmodes (the
Galerkin approximation), which accurately reproduces the full dynamics, and
then analyze the system's dynamical spectrum. The analysis enables us to
interpret the observed quasi-integrability as the fact that the finite-mode
dynamics always produces a \emph{quasi-discrete} power spectrum, with no
visible continuous component, the presence of the latter being a necessary
manifestation of ergodicity. This conclusion remains true when a strong
random-field component is added to the initial conditions. On the other
hand, the same analysis for the GPE in an infinitely deep potential box
leads to a clearly continuous power spectrum, typical for ergodic dynamics.
\end{abstract}

\vspace{2pc} \noindent\textit{Keywords}: dark soliton, integrability,
Gross-Pitaevskii equation, sound waves, phonons, Galerkin approximation,
Bose-Einstein condensate

\maketitle

%\pacs{00.00, 20.00, 42.10}

\section{Introduction}

Integrability, relaxation, and thermalization of many-body systems are
intricately-linked key topics of the modern theory of non-equilibrium
dynamical systems. Although, strictly speaking, a closed quantum system
should exhibit no thermalization in the usual sense, non-integrable closed
systems can nonetheless mimic relaxation to thermal equilibrium through
dephasing occurring within the eigenstate thermalization hypothesis \cite%
{srednicki_1994,rigol_2008}.

The investigation of these issues has recently become a core activity in
studies of dynamics of ultracold gases \cite{langen_2015}, due to the
uniquely precise experimental control achieved in this field. Such settings
can be engineered in both weak- and strong-interaction regimes, the
effectively one-dimensional (1D) realizations being of particular relevance,
as the respective model equations may be able to support integrable
dynamics. In this context, pioneering experiments with ultracold atoms in
the effectively 1D regime have revealed evidence for a long-term absence of
thermalization \cite{kinoshita_2006}, attributed to the expected
integrability of the underlying (Lieb-Liniger) model of strong interactions.
Subsequent works, however, have predicted timescales for the breakdown of
integrability in experimentally relevant geometries, with thermalization
possible through virtual excitation of higher radial modes \cite%
{mazets_2008,mazets_2010}. These findings were reported to be consistent
with both the previous experiments \cite{kinoshita_2006} and other relevant
observations \cite{hofferberth}, subsequent work also addressing the
emergence of pre-thermalization \cite{prethermalization}, in which a closed
system loses part of its initial information.

In the idealised setting described by an integrable equation, which
possesses an infinite number of conserved quantities, the trajectories are
weakly sensitive to initial conditions, lying on invariant tori in the phase
space, realistic systems often exhibit \textquotedblleft weak integrability
breaking", in the sense that one can construct and probe \textquotedblleft
quasi-conserved" quantities. One should here distinguish between two
different issues: the perceived presence of (quasi-)integrability of a given
physical system as probed in experiments, and the emergence of integrability
in the equations believed to accurately describe the physical system, which
is usually probed through numerical simulations. The fundamental equation
describing ultracold atoms in the weakly-interacting regime is the nonlinear
Schr\"{o}dinger equation (NLSE), with cubic nonlinearity arising from
inter-atomic collisions, alias the Gross-Pitaevskii equation (GPE). This
equation is the workhorse of the theoretical studies of ultracold atoms,
with an impressive portfolio of successes in predicting experimental
phenomena to high accuracy, including the static characteristics of the
ultracold gases, their modes, nonlinear waves, dynamical instabilities, etc.
\cite{pethick, BEC,barenghi}. The most common case to which we limit our
study here is when the effective interactions are repulsive (i.e., the
respective nonlinearity is defocusing). Such an equation is known to be
integrable in the 1D free space (including the case of periodic boundary
conditions) \cite{zakharov_1973,kichenassamy_1996,ma_1981,dodd_1982}, but
not in the presence of the harmonic-oscillator confining potential, which is
relevant for modeling actual experiments. Even in this case, however,
long-time simulations of the 1D GPE have revealed no conclusive evidence of
chaotization \cite{cockburn_2011,mazets_2011}, which is believed to
originate in the experiment from the coupling to transverse degrees of
freedom, beyond the limits of the 1D approximation \cite{mazets_2010}. On
the other hand, a single particle in the harmonic-oscillator trap is
commonly known to be integrable. The question then arises under what
conditions, and to what extent, features of the integrability may be
approximately preserved in many-body systems trapped by this potential.

The closest many-body state which exhibits some particle-like properties is
a solitonic excitation -- specifically, a dark soliton in the case of
repulsive interactions, which is thus a natural candidate to use as a probe
of the integrability. Importantly, previous studies of the motion of dark
solitons in the harmonic-oscillator-trapped 1D GPE lead to a quasi-periodic
evolution, revealing no evidence of chaotization (ergodicity) in the
evolution of the mean-field wavefunction, unlike certainly non-integrable
settings, corresponding to other (anharmonic) probed trapping potentials
\cite{Parker2004,parker_2010}. This observation suggests an
\textquotedblleft apparent quasi-integrability" of the 1D GPE in the
harmonic-oscillator trap, with regard to the motion of a dark soliton, which
was predicted to perform shuttle motion, as a classical particle, with a
well-defined oscillation amplitude and frequency \cite%
{fedichev_1999,Muryshev1999,busch_2000,huang_2002,Theocharis2007,frantzeskakis_2010,parker_2003}%
. This behavior is consistent with experiments which have generated dark
solitons and demonstrated their motion in elongated quasi-1D BECs \cite%
{DS_expts,weller_2008}. However, the presence of any potential, \emph{%
including} the harmonic-oscillator trap, is known to break the integrability
of the underlying GPE, and, in particular, to trigger the emission of
small-amplitude excitations (\textquotedblleft sound waves") from dark
solitons moving with acceleration \cite{busch_2000,huang_2002,
parker_2003,Parker2004,pelinovsky_2005,allen_2011,radouani_2003}. This
mechanism of the decay of dark solitons into radiation is similar to that
known for optical dark solitons governed by the NLSE \cite%
{Pelinovsky1996,kivshar_1998,proukakis_2004_2}.

To reconcile these apparently contrasting predictions, one implying the
presence of the effective quasi-integrability, and the other referring to
the non-integrability of the GPE with the harmonic-oscillator potential, it
was proposed that the emission of sound waves might be reversible, i.e.,
that the dark soliton may reabsorb the emitted waves, thus stabilizing
itself against the systematic decay \cite{parker_2010,parker_2003,allen_2011}%
. This effect may even be employed to preferentially stabilize dark solitons
in states with selected energies \cite{proukakis_2004}. The reversibility
effect has been shown to be crucial over timescales shorter than those
imposed by other non-integrability factors (for example, those related to
thermal dissipation and coupling to the transverse dimensions) in
harmonic-oscillator-trapped BECs \cite%
{fedichev_1999,Muryshev1999,proukakis_2004,timescales}. In turn, the
sound-emission reversibility suggests that the harmonic-oscillator potential
may maintain quasi-integrability of the system. Further evidence to support
this conjecture comes from simulations which reveal a systematic decay when
the harmonic-oscillator potential is altered, and the quasi-integrability is
clearly broken, e.g., by the addition of dimple traps \cite%
{busch_2000,parker_2003}, an optical lattice \cite{Parker2004,Kevrekidis2003}%
, or a localized obstacle \cite{Frantzeskakis2002,Proukakis2004}. Another
sign of the quasi-integrability in the presence of the harmonic-oscillator
confinement is an essentially elastic character of collisions between two
trapped dark solitons, observed in direct simulations and verified
experimentally \cite{weller_2008,theocharis_2010}.

%Further, the simulations
%demonstrate that emission of small-amplitude excitations
%(acoustic waves") by a dark soliton (DS), performing shuttle motion in the harmonic-oscillator
%trap (which is a major manifestation of the \ nonintegrability of the
%underlying GPE), is reversible, i.e., the same DS may reabsorb the emitted
%waves, which prevents systematic its systematic decay \cite{busch_2000,parker_2003,pelinovsky_2005,parker_2010,allen_2011}. The reversibility effect has been shown to be crucial over
%timescales shorter than those imposed by other non-integrability effects in
%harmonic-oscillator-trapped Bose-Einstein condensates (BECs) \cite{timescales}. Another sign
%of the quasi-integrability is an essentially elastic character of collisions
%between two trapped DSs, observed in direct simulations \cite{weller_2008,theocharis_2010}.

To gain insight into the presumably quasi-integrable dynamics, we here
develop a finite-mode approximation for the 1D GPE with the
harmonic-oscillator potential, known as the \textit{Galerkin approximation}~%
\cite{GA_paper}: the wave field is expanded over the full set of eigenmodes
of the linear Schr\"{o}dinger equation with the harmonic-oscillator
potential, thus replacing the underlying cubic GPE by a chain of nonlinearly
coupled ordinary differential equations for the evolution of amplitudes of
the eigenmode expansion. The chain is truncated for a finite set of $M$
modes, sufficient to provide an accurate approximation for the global
evolution of the mean-field wave function governed by the GPE, including
relevant features such as the above-mentioned sound emission and absorption
by the dark soliton. A similar expansion approach was developed for various
nonlinear models \cite{Galerkin approximation-applications}, including
multi-component and multi-dimensional GPE systems \cite{Radik,Chinese}.

The finite-mode Galerkin expansion is also at the heart of the projected
Gross-Pitaevskii equation (PGPE) \cite{blakie_2007}, which has been
extensively applied to model Bose gases at finite temperatures. However,
there are several contextual differences between our study and those using
the PGPE. Specifically, we seek to approximate the zero-temperature GPE wave
field, not a thermal field, with the key point being that we can very
accurately capture the soliton dynamics and aspects of quasi-integrability
by employing $M=16$ modes.

The aim of our analysis, performed in the framework of the suitably
truncated finite-mode dynamical system, is to highlight the degree of the
quasi-integrability of the underlying cubic GPE including the
harmonic-oscillator potential. Specifically, we find, with high numerical
accuracy, that the power spectrum of all dynamical trajectories remains
\emph{quasi-discrete} in the course of the indefinitely long evolution,{\
corresponding to a quasi-periodic motion, rather than to chaotic dynamics.
This observation strongly suggests that the Galerkin-approximation system
with a finite number of the degrees of freedom has almost all its
trajectories spanning invariant tori, in accordance with the
Kolmogorov-Arnold-Moser theorem \cite{KAM,KAM2}. Such a finding provides an
adequate explanation of the effective quasi-integrability featured by the
underlying harmonically-trapped GPE in the previously reported direct
simulations \cite{parker_2003}. }

%In this work, we present results for the GPE with different numbers of modes, up to $M=16$.

% First, the comparison of the full GPE simulations with the results produced by the truncated Galerkin approximation demonstrates that it indeed provides a very accurate approximation for the evolution of the mean-field wave function. Keeping up to $M=16$ modes in the approximation is sufficient to reproduce in a virtually exact form, for indefinitely long times, such basic dynamicalregimes as oscillations of a DS in the trap %, collision between two DSs, and emission and re-absorption of sound waves by the DS. In addressing these settings, we use, as the control parameter, the total norm, $N$, of the mean-field wave function, while the scaled nonlinearity coefficient in the GPE and the strength of the harmonic-oscillator potential are scaled to be $1$. Naturally, for very large values of $N$, a larger set of modes, which are defined in terms of the linear limit, is needed to correctly approximate the strongly nonlinear configuration. In particular, as $N$ increases by a fac
% tor of $100$%, the number of modes necessary for constructing the accurate Galerkin approximation grows from $% 4$ to $12$. The Galerkin approximation with $M=16$ turns out to be sufficient for the virtually exact representation of all the full GPE\ solutions.

Although the analysis reveals strong evidence of the repeated reversible
cycles of the emission/absorption of radiation from/by the dark soliton,
there is no straightforward way to isolate the soliton and sound modes
through the Galerkin approximation in the condensate trapped in the
harmonic-oscillator potential. To demonstrate the role of this process in a
more explicit form, we also develop a similar analysis for the GPE in a
potential box with zero boundary conditions (i.e., an infinitely deep
rectangular potential, which can be experimentally realized using
electromagnetic fields \cite{box_traps}, although the box walls in the
experiment are softer than the ideal impenetrable ones). The Galerkin
approximation for the potential box can be naturally built on the basis of
the underlying sine and cosine eigenfunctions \cite{students}. Systematic
simulations of the GPE in the infinitely deep box show that a
moving dark soliton shuttles back and forth in a stable manner (although the
soft walls may cause an instability and sound emission \cite{sciacca}).
Interestingly, and perhaps somewhat unexpectedly, we find in this case that
the power spectrum of generic trajectories is continuous, in direct contrast
to the quasi-discrete spectrum found in the harmonic-oscillator potential,
which clearly suggests chaotization (ergodicity) of the dynamics in the box,
rather than evolution guided by invariant tori. Such behaviour is also
wholly captured by our finite-mode expansion (without the need for using the
known exact box eigenstates of the nonlinear equation \cite{carr}).

It is relevant to mention that, strictly speaking, the NLSE in a finite
interval with zero boundary conditions belongs to the class of integrable
equations \cite{Maxim}. This fact seems to be in contradiction with the
above-mentioned ergodicity revealed by the Galerkin approximation for the
potential box. However, it is known that there are two types of
integrability, \textit{strong} and \textit{weak} \cite{Zakharov}. The
contradistinction between them is based on the relation between the numbers
of degrees of freedom and dynamical invariants. Indeed, a
Liouville-integrable dynamical system with a finite number of degrees of
freedom must have it equal to the number of dynamical invariants \cite%
{Liouville}. In the limit of the infinite number of degrees of freedom
(integrable PDEs), the set of dynamical invariants is also infinite, but in
the case of weak integrability the set is incomplete (not \textquotedblleft
sufficiently infinite"), which allows the system to feature non-integrable
dynamics, such as fission and merger of solitons in the weakly integrable
three-wave system \cite{three}, another known example of weak integrability
being provided by the Kadomtsev-Petviashvili-I equation. Models of this
type, in spite of their formal integrability, readily admit chaotic dynamics
-- in particular, in the form of wave turbulence in the free space \cite%
{Zakharov}. Of course, the finite-mode truncation most plausibly breaks the
strong and weak integrability of the underlying partial differential equation
alike, but the concept of the weak integrability suggests a possible
explanation to the fact that the truncation, derived for a weakly integrable
model, may feature ergodicity: if the underlying model admits chaotic
dynamics, the truncated version may feature it too. Concerning the GPE in
the finite-size box, the issue of its strong/weak integrability is not
explored yet, to the best of our knowledge. This issue may be a subject for
a separate study, which is definitely beyond the scope of the present work.

The rest of the paper is organized as follows. In Section II, we summarize
the Galerkin approximation for both the harmonic-oscillator and box traps
and demonstrate its success in capturing both the ground-state solutions and
dark-soliton motion, with only a small number of modes in the truncation
(technical details are given in Appendices A and B). This finding enables us
to use the motion of the dark soliton as a probe for the quasi-integrability
of the 1D harmonically-confined GPE, focusing in Section III on the
distinction between the quasi-discrete and continuous spectra of the
evolution of complex amplitudes of the Galerkin truncation, which are found,
respectively, in the harmonic-oscillator and box traps. Our findings are
summarized in Section IV.

% Our support for the notion of the quasi-integrability under the harmonic-oscillator confinement is presented in Section III, based on the distinction between the discrete and a quasi-continuous spectra of the evolution of Galerkin approximation\ complex amplitudes. The paper is concluded by Section IV.

%The elaboration of the Galerkin approximation with a sufficiently large number of modes, $M$, is practically possible in the infinite hard box case, as the eigenmodes of the respective linear Schr\"{o}dinger equation are commonly known in the form of cosines and sines. The result of the corresponding analysis is: the Galerkin approximation with relatively large $M$ (in particular, again with $M=16$) also provides a very accurate approximation for the respective GPE numerical solutions, but the agreement eventually deteriorates. As concerns the dynamics of the finite-mode system per se, which corresponds to the box potential, the power spectrum of generic trajectories turns out to be \emph{continuous}, on the contrary to the above-mentioned quasi-discrete spectrum in the case of the harmonic-oscillator potential, which clearly suggests chaotization (ergodicity) of the dynamics, rather than evolution guided by invariant tori.

%\section{The analytical considerations and their verification}

\section{The Galerkin approximation and its validity}

Our analysis starts from the well-known 1D GPE, written in the presence of
an arbitrary time-independent potential, $V(x)$ \cite{pethick,BEC},
\begin{equation}
i\hbar \frac{\partial \Psi }{\partial t}=-\frac{\hbar ^{2}}{2m}\frac{%
\partial ^{2}\Psi }{\partial x^{2}}+{g}|\Psi |^{2}\Psi +V(x)\Psi .
\label{eqn:gpe}
\end{equation}%
%
%
%
%
%
%
%where $\int \mathrm{d}x |\Psi(x,t)|^2=\bar{N}(t)$ gives the particle number at time $t$.
Here $\Psi (x,t)$ is the mean-field wave function of the BEC, normalised to
the number of particles $\mathcal{N}=\int |\Psi |^{2}\,\mathrm{d}x$, and $g$
is the coefficient of the cubic nonlinearity, induced by the van der Waals
interactions between atoms which make up the BEC. The characteristic energy,
length and time scales are the chemical potential $\mu $ of the BEC, the
healing length $\xi =\hbar /\sqrt{m\mu }$, and $\tau =\xi /c$, where $c=%
\sqrt{\mu /m}$ is the speed of sound, and $m$ is the atomic mass. Using
these scales to define dimensionless energy, position and time variables,
Eq.~(\ref{eqn:gpe}) can be rewritten as
\begin{equation}
i\frac{\partial \psi }{\partial \tilde{t}}=-\frac{1}{2}\frac{\partial
^{2}\psi }{\partial \tilde{x}^{2}}+\sigma |\psi |^{2}\psi +\tilde{V}(\tilde{x%
})\psi ,  \label{GPE}
\end{equation}%
where $\psi=\sqrt{2|a_s|/l_x^2}\Psi$ and $\sigma =+1$ and $-1$ corresponds
to the repulsive and attractive nonlinearities, respectively. The
dimensionless wave function $\psi \equiv \psi (\tilde{x},\tilde{t})$ is
subject to normalization $N=\int |\psi |^{2}\,\mathrm{d}\tilde{x}$, where $%
\mathcal{N}=\mu\xi N/g$. Thus, for typical experimental parameters the atom
number will correspond to $\mathcal{N}\sim10^4N$. From this point on, we
drop the tilde notation for dimensionless variables; the exception is in
figures, where $x$ and $t$ are presented in the dimensional form.

In this work, we are interested in repulsive interactions which admit dark
solitons trapped in the external potential \cite{busch_2000}, therefore we
fix $\sigma =1$. We consider harmonic-oscillator and box potentials, which
are defined, respectively, as
\begin{equation}
V(x)=\frac{1}{2}\omega _{x}^{2}x^{2},~\omega _{x}\equiv 1,~~\mathrm{or}~\ \
V(x)=\left\{
\begin{array}{c}
0,~\mathrm{at~}~0<x<L, \\
\infty \,,~~\mathrm{elsewhere.~}%
\end{array}%
\right.  \label{pot}
\end{equation}%
In the former case, $\omega _{x}\equiv 1$ is fixed by rescaling. It is
relevant to note that the infinite-box potential, which gives rise to zero
boundary conditions, $\psi (x=0)=\psi (x=L)=0$, directly applies, in
addition to BEC, as the model of a metallic conduit for microwaves \cite%
{micro}.

The Galerkin approximation takes two different forms, depending on the
potential considered. In each case, the wave field is approximated by an $M$%
-mode linear combination of time-dependent eigenmodes of the corresponding
linear Schr\"{o}dinger equation, with each eigenmode subject to the unitary
normalization. In the harmonic-oscillator case, the corresponding ansatz is
\begin{equation}
\psi _{\mathrm{GA}}(x,t)=\sum_{n=0}^{M-1}a_{n}(t)\exp \left( -\frac{x^{2}}{2}%
-i\left( \frac{1}{2}+n\right) t\right) \frac{H_{n}(x)}{\pi ^{1/4}\sqrt{%
2^{n}n!}},  \label{ga_harm}
\end{equation}%
where $a_{n}(t)$ are complex amplitudes, which are slowly varying functions
of time, in comparison with $\exp \left( -i\left( \frac{1}{2}+n\right)
t\right) $, and $H_{n}(x)$ are the Hermite polynomials. In the case of the
box potential, the expansion is built as
\begin{equation}
\psi _{\mathrm{GA}}(x,t)=\sum_{n=0}^{M-1}a_{n}(t)\sqrt{\frac{2}{L}}\sin
\left( \frac{\left( n+1\right) \pi x}{L}\right) e^{-iE_{n}t},
\label{eqn:ga_box}
\end{equation}%
with $E_{n}=\pi ^{2}\left( n+1\right) ^{2}/2L^{2}$, and amplitudes $a_{n}(t)$
being slowly varying functions in comparison with $\exp \left(
-iE_{n}t\right) $.

Note that it would also be possible to conduct the expansion using the
nonlinear modes of the system. However, while the nonlinear solutions in the
1D box potential are known \cite{carr}, they are not known in analytical
form for the harmonic potential. Moreover, the nonlinear box solutions would
give rise to extremely cumbersome differential equations. Hence for the
purpose of this work we focus on expansion into linear modes, which are
clearly demonstrated to capture the key features of the underlying GPE they
are approximating.

Evolution equations for amplitudes $a_{n}$ can be readily derived by means
of the variational principle \cite{Anderson,Progress}. To this end, we use
the Lagrangian of Eq.~(\ref{GPE}),
\begin{equation}
L=\int_{-\infty }^{+\infty }\left( i\psi ^{\ast }\frac{\partial \psi }{%
\partial t}-\frac{1}{2}\left\vert \frac{\partial \psi }{\partial x}%
\right\vert ^{2}-\frac{\sigma }{2}|\psi |^{4}-V(x)|\psi |^{2}\right) \mathrm{%
d}x.  \label{L}
\end{equation}%
The substitution of \textit{ans\"{a}tze} defined by Eqs.~(\ref{ga_harm}) and
(\ref{eqn:ga_box}) into the Lagrangian leads to the following result,
\begin{equation}
L=i\sum_{n=0}^{M-1}a_{n}^{\ast }\frac{da_{n}}{dt}-H,  \label{L-H}
\end{equation}%
where the Hamiltonian is
\begin{equation}
H=f(a_{0},...,a_{M-1},a_{0}^{\ast },...,a_{M-1}^{\ast },t),  \label{eq:H}
\end{equation}%
and function $f$ is a combination of quartic terms, depending on the number
of modes kept in the Galerkin approximation. Accordingly, the dynamics are
governed by the Euler-Lagrange equations derived from the Lagrangian,%
\begin{equation}
\frac{da_{n}}{dt}=-i\frac{\partial H}{\partial a_{n}^{\ast }}.  \label{EL}
\end{equation}
This is a mechanical system with $M$ degrees of freedom and two dynamical
invariants, $H$ and the total norm,
\begin{equation}
N=\sum_{n=0}^{M-1}\left\vert a_{n}\right\vert ^{2}.  \label{eqn:N}
\end{equation}%
An explicit form of the Hamiltonian and dynamical equations for the Galerkin
approximation with $M=4$ are given in Appendix A. Similar equations have
been explicitly derived up to $M=16$ (they are not included here, as they
seem too cumbersome, but, nevertheless, they are tractable, for the purposes
of the current analysis.)
%Note that the Projected GPE method \cite{blakie_2007} is based on such a decomposition over a large number of modes, but such numerics has typically been focussed on higher-dimensional settings.

We employed a Crank-Nicolson method to find stationary and dynamic solutions
to the GPE in MATLAB. Typical simulation parameters are (in the scaled
units): spatial discretization $\Delta x=0.05$, simulation-box length $L=20$%
, and time step $\Delta t=0.001$. The finite-mode dynamical system based on
Eq.~(\ref{EL}) was also solved with the help of MATLAB. The largest mode
number considered in this work is $M=16$, the consideration of still larger $%
M$ being technically possible, but not really necessary, as shown by the
results presented below. The MATLAB codes used to solve Eq.~(\ref{EL}) for $%
M=4,8$ and $16$, in the presence of the harmonic-oscillator and box
potentials, are available online \cite{data}.

In Appendix B we evaluate accuracy of the Galerkin approximation in
capturing the ground state of the system. In the case of the
harmonic-oscillator trap and $M=16$, the agreement is almost perfect for all
norms considered. The agreement is almost as good for the box trap: for $%
M=16 $, nearly perfect agreement is obtained for the case of low norm $N\le10
$. Similarly, with $M=16$ modes, the truncated Galerkin approximation is
sufficient to reproduce in a virtually exact form the evolution of the
mean-field wave function in the presence of harmonic confinement over
indefinitely long times, thus accurately reproducing the oscillations of a
dark soliton in the trap which fully incorporates the
previously-characterised emission and re-absorption of sound waves by the
dark soliton. Note that, contrary to the harmonic-oscillator model, the
agreement for the infinitely deep box does eventually deteriorate, but only
in the course of very long time evolution.

%As we show below, the comparison of the full GPE simulations with the results produced by the truncated Galerkin approximation demonstrates that it indeed provides a very accurate approximation for the evolution of the mean-field wave function. Keeping up to $M=16$ modes in the approximation is sufficient to reproduce in a virtually exact form, over indefinitely long times, such basic dynamical regimes as oscillations of a dark soliton in the trap and the emission and re-absorption of sound waves by the dark soliton. In addressing these settings, we use, as a control parameter, the total norm, $N $, of the mean-field wave function. Naturally, for very large values of $N$, a larger set of the expansion modes, which are defined in terms of the linear limit, is needed to correctly approximate the strongly nonlinear configuration. In the case of the infinitely deep box, the elaboration of the Galerkin approximation with a sufficiently large number of modes, $M$, is also practica
% lly possible, the detailed analysis showing that $M=16$ also provides a very accurate approximation for numerical solutions of the respective GPE, but the agreement eventually deteriorates in the course of very long evolution, on the contrary to the harmonic-oscillator model.

\subsection{Dark-soliton dynamics as the testbed for the validity of the
Galerkin approximation}

Next we test the validity of the Galerkin approximation for the dark soliton
through its comparison to the numerically exact GPE solution. In the case of
the repulsive nonlinearity ($\sigma >0$), which we deal with in this work,
the free-space GPE (no trap) has a commonly known family of dark-soliton
solutions, written here in the unscaled units \cite{zakharov_1973},
\begin{equation}
\psi _{\mathrm{ds}}(x,t)=\sqrt{n_{0}}\left[ \beta \tanh \left( \frac{%
x-x_{0}+vt}{\xi }\beta \right) +i\left( \frac{v}{c}\right) \right] e^{-i\mu
t/\hbar }.  \label{DS}
\end{equation}%
Here $\beta =\sqrt{1-v^{2}/c^{2}}$, $x_{0}$ is the initial position and $v$
the soliton's velocity. Stationary (alias black) solitons, with $v=0$, have
a zero-density notch with a phase slip of $\pi $ across it. The soliton's
energy decreases with increasing speed \cite{kivshar_1998}, emulating a
particle with a negative effective mass \cite{busch_2000}.
\begin{figure}[b]
\centering
\hspace{-1cm} Harmonic trap \hspace{4.5cm} Box trap
\par
\includegraphics[width=0.45\columnwidth]{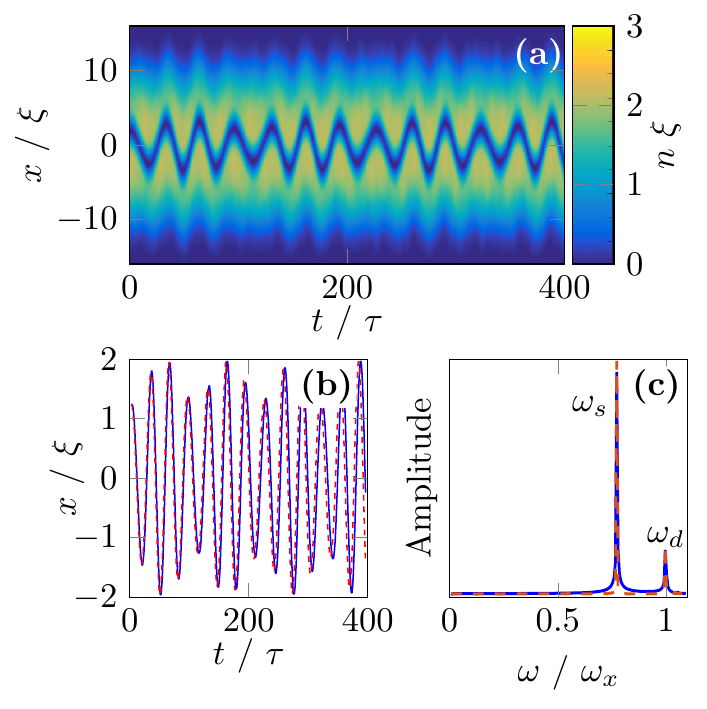} %
\includegraphics[width=0.45\columnwidth]{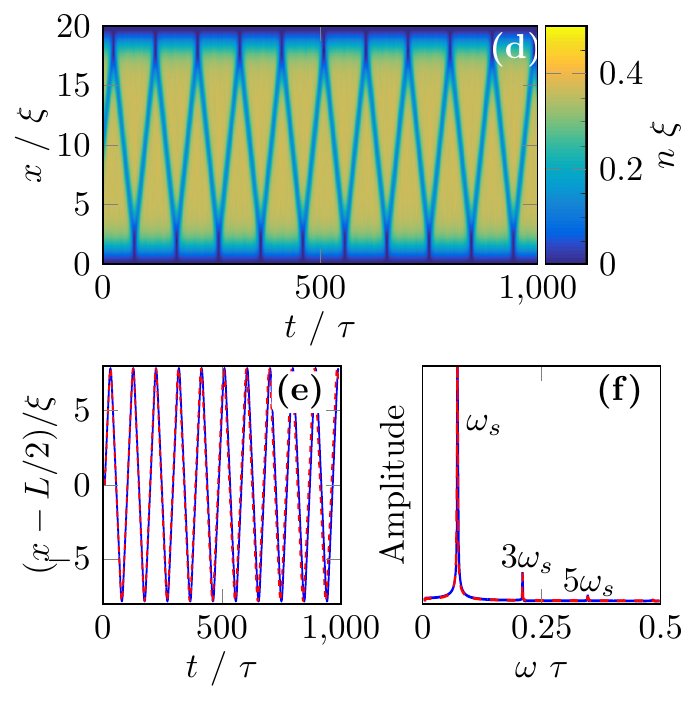}
\caption{The motion of an initially off-center black soliton ($v=0$) in the
harmonic-oscillator (left column) or gray soliton ($v=0.6$) in the box
(right column) traps. Panel (a) depicts the evolution of the complex
amplitudes as per the Galerkin approximation for the ansatz with $16$ modes,
with similar results for the box trap shown in (d). Panels (b) and (e)
display the corresponding center-of-mass oscillations of the dark soliton,
as produced by simulations of the GPE (blue solid lines) and by the Galerkin
approximation (red dashed lines), while their Fourier transforms are shown
in (c) and (f). Parameters are $N=15$, $v=0$ and $x_{0}=1$ (see Eq.~(\protect
\ref{DS})) for the harmonic-oscillator trap; $N=5$, $L=20$, $v=0.6$ and $%
x_{0}=10$ for the box.}
\label{fig:osc_eg}
\end{figure}

Figure \ref{fig:osc_eg} shows oscillations of a dark soliton in the
harmonic-oscillator and box trapping potentials (left and right columns,
respectively). The first row displays the evolution predicted by the
Galerkin approximation, as produced by the solution of Eq.~(\ref{EL}), with $%
M=16$ modes. In the box trapping potential, the soliton trajectory closely approximates a triangular wave, as would be expected.  In the harmonic-oscillator potential, one might expect that the soliton would closely follows a sinusoidal trajectory; however, the interaction of the soliton with the dynamical background condensate significantly distorts the trajectory. The deformation of the background field
reveals evidence for the interaction of the soliton with the sound
(propagating excitations), whereas, in the potential box, we observe
chaotization of the soliton dynamics at very long evolution times.

% In the case of the harmonic-oscillator confinement the deformation of the background field shows the interaction of the soliton with the sound (propagating excitations), as produced by the Galerkin approximation. Similar simulations with smaller mode number $M=4$ have also been performed (not shown here in detail). In harmonic confinement the soliton  performs sinusoidal oscillations buffeted by violent oscillations of the background condensate, whereas in the box potential the soliton motion shows evidence of chaotization, making it impossible to separate it from the background evolution. The dark soliton's chaotization is eventually realized too in the same box at longer times for $M=8,16$.}

It is relevant to mention that the above-mentioned oscillations of the
background condensate trapped in the harmonic-oscillator potential (panel
(a)) are excited by the sound emission from the accelerating soliton. The
wavelength of the sound is comparable to the system size, hence it becomes
visible as a dipole oscillation of the cloud (note that the direct
visualisation of the sound pulse is challenging due to the immediate
reflection of the sound from boundaries and re-interaction with the soliton,
motivating the use of dimple traps elsewhere to overcome this issue \cite%
{parker_2010,parker_2003}). A quasi-steady state is thus established,
wherein the background modes and soliton maintain constant average
amplitude. This equilibrium is attributed to the balance between emission
and reabsorption of sound by the soliton \cite{parker_2010,parker_2003}.
Fluctuations in this energy balance are evident in the quasi-periodic
acceleration and deceleration of the soliton, visible in the panel (a).

%
%It is relevant to mention that the above-mentioned reversible emission and
%reabsorption of acoustic waves by the dark soliton performing shuttle motion
%in the harmonic-oscillator potential can be observed in panels (a)-(c) of
%Fig. \ref{fig:osc_eg}. Indeed, the dark soliton becomes shallower and gains
%speed after emitting an acoustic wave packet, allowing for a larger
%amplitude of shuttle oscillations. The soliton becomes slower and deeper,
%after reabsorbing the sound. It is seen in the figure that the emission and
%reabsorption occur quasi-periodically, while the dark soliton remains a
%robust mode.

Figures \ref{fig:osc_eg}(b) and (e) display the soliton's center-of-mass
motion, with overlaid results produced by the predictions of the Galerkin
approximation with $M=16$ and GPE\ simulations.
%\tom{ After ten oscillations, splitting between the two trajectories is less than $0.1$ for each potential: IN WHAT UNITS? NOT SURE WHAT THIS MEANS HERE}.
The Fourier transform of these center-of-mass oscillations is displayed in
Figs.~\ref{fig:osc_eg}(c) and (f), where the oscillation frequency of the
soliton, $\omega _{s}$, is highlighted, along with frequency $\omega _{d}$ \
[in panel (b)] corresponding to the dipole mode of small excitations of the
condensate as a whole. The latter mode, with%
\begin{equation}
\omega _{d}=\omega _{x}\equiv 1,  \label{ds}
\end{equation}%
[see Eq.~(\ref{pot}) for the definition of $\omega _{x}$], is excited by the
motion and sound emission of the dark soliton traversing the condensate \cite%
{parker_2003}.

The dark soliton in the harmonic-oscillator potential is known to oscillate
at frequency
\begin{equation}
\omega _{s}=\omega _{x}/\sqrt{2},  \label{sx}
\end{equation}%
as shown theoretically \cite{busch_2000} and experimentally \cite%
{weller_2008}, deep in the Thomas-Fermi limit, corresponding to large $N$ in
our notation. The role of the total number of modes, $M$, of the Galerkin
approximation is addressed in Appendix B, demonstrating perfect dynamical
accuracy of the approximation for $M=16$ modes. Moreover, the dark soliton's
oscillation frequency in the harmonic-oscillator potential indeed
approaches, as expected, the value $\omega _{x}/\sqrt{2}$ as $N$ increases.

\section{Probing quasi-integrability in the 1D harmonic-oscillator potential}

%As said above, the main incentive for undertaking the present analysis is the quasi-integrability, suggested by previous simulations of the one-dimensional GPE under the harmonic-oscillator confinement. In the absence of the potential, the GPE reduces to the nonlinear Schr\"{o}dinger equation, whose integrability in one dimension is commonly known
%\cite%{zakharov_1973,kichenassamy_1996, dodd_1982}.

\begin{figure}[tbp]
\centering
\hspace{0.0cm} Harmonic trap \hspace{4.5cm} Box trap
\par
\includegraphics[scale=1.1]{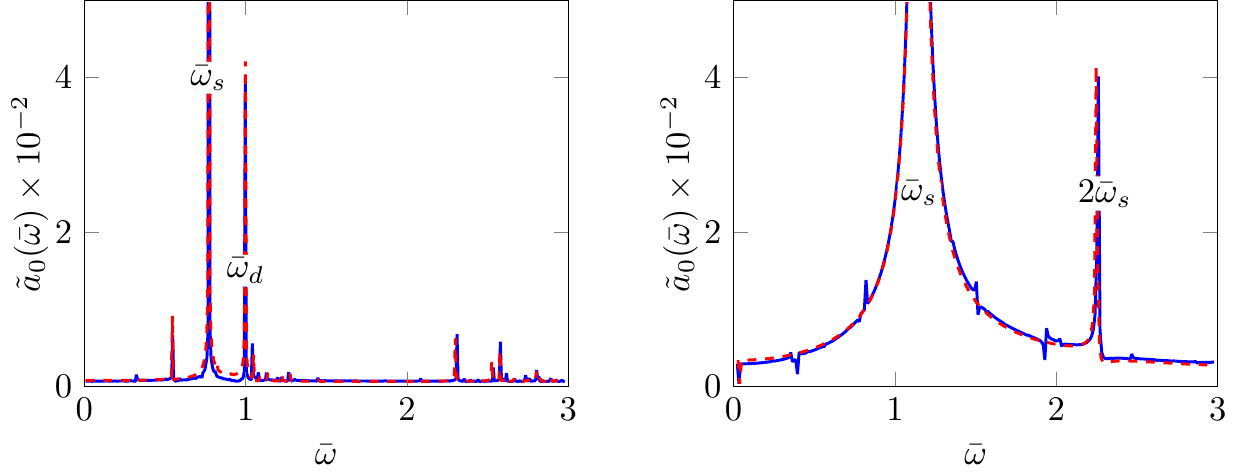}\newline
\includegraphics[scale=1.1]{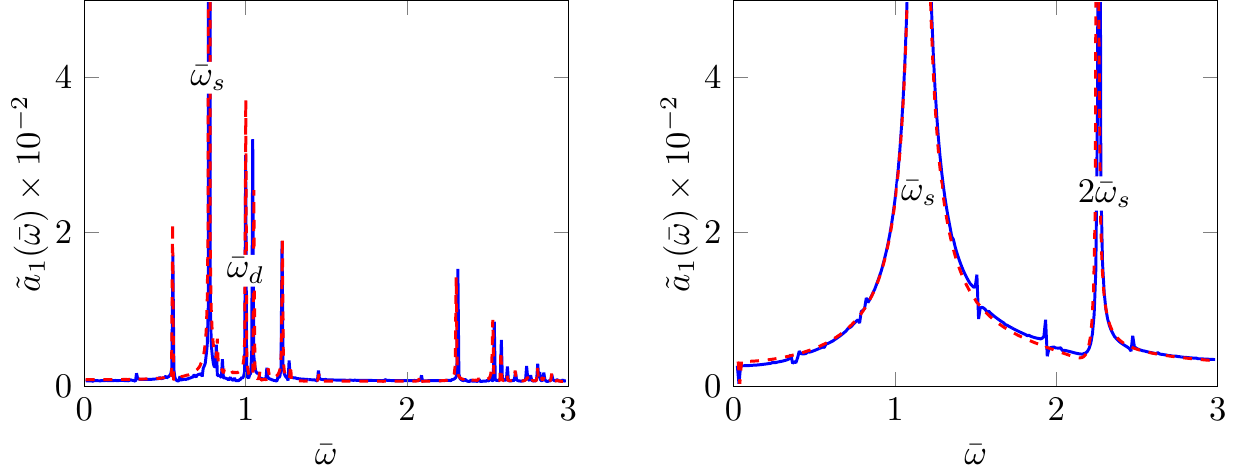}\newline
\caption{ Generic examples of power spectra for the first (top row) and
second (bottom row) amplitudes of the Galerkin expansion in the models with
the harmonic-oscillator (left) and box (right) potentials. Parameters are
the same as in Fig.~\protect\ref{fig:osc_eg} with $M=16$. In each case, the
simulations comprised $100$ full periods of shuttle oscillations of the dark
soliton. As in Figs. \protect\ref{fig:osc_eg}(b,c,e,f), we show results
produced by both the GPE simulations (solid blue lines) and Galerkin
approximation with $16$ modes (red dashed lines), which reveals excellent
agreement between both.}
\label{fig:integ}
\end{figure}

In this section, we address the challenging issue of detecting
quasi-integrability of the GPE with the harmonic-oscillator potential, which
is the main reason why the above analysis was undertaken. As is well known,
in strictly integrable dynamical systems the power spectrum of the time
dependence of dynamical variables (the complex amplitudes, in the present
case), $\tilde{a}_{j}(\omega )=\mathcal{F}[|a_{j}(t)|^{2}]$, where $\mathcal{%
F}$ stands for the Fourier transform, is truly discrete, corresponding to
the generic quasi-periodic motion on a surface of an invariant torus, while
non-integrable systems feature a conspicuous continuous component in the
spectrum, as a result of destruction of the tori \cite{KAM,Moon}. We have
applied this criterion of the integrability, by analysing the motion
displayed in Fig.~\ref{fig:osc_eg}, extending the computation of the spectra
to a hundred oscillations of the dark soliton.
%, \tom{keeping the simulation time small enough to minimize any growing
%numerical error}.
Figure \ref{fig:integ} depicts our main findings, both for the evidently
discrete spectrum in the harmonic-oscillator trap, and the case of the box
potential (left and right images, respectively). The results are represented
by power spectra $\tilde{a}_{0}(\omega )$ [top panels] and $\tilde{a}%
_{1}(\omega )$ [bottom panels] of the first and second amplitudes of the
Galerkin expansion, which are overlaid on the corresponding results of the
GPE\ simulations. The GPE spectra were produced by computing the
corresponding amplitudes as per Eq.~(\ref{fid}), and then calculating their
power spectra. In both cases, the agreement between the Galerkin
approximation and full GPE simulations is impressive, a feature which is
also true for higher-order $\tilde{a}_{j}(\omega)$ coefficients (not shown
here).

A crucial finding is the stark difference in the results for the dark
soliton's motion in the two potentials: In the harmonic-oscillator case
(left) we obtain a spectrum consisting of extremely sharp peaks, which may
be definitely categorized as a \emph{practically discrete} spectrum, thus
representing \emph{quasi-integrable dynamics}. The tallest peaks in the
spectrum can be immediately identified as located at the above-mentioned
frequencies $\omega _{d}$ and $\omega _{s}$ of the dipole mode of the
excitations of the condensate as a whole [see Eq.~(\ref{sx})], and shuttle
oscillations of the dark soliton [see Eq.~(\ref{ds})]. The surrounding peaks
can be readily identified as combinational frequencies produced by mixing of
these two modes. In stark contrast to this, the spectrum in the box trap
exhibits a broad peak, which clearly represents a continuous spectrum,
typical to non-integrable systems, that give rise to dynamical chaos and
ergodicity \cite{Moon}. Simulations of the dynamical Galerkin system for the
potential-box model with $M<16$ produce a similar behaviour, but with a
growing noise component.

%In the case of the box trap, the spectrum exhibits a broad peak, virtually identical (as well as the entire spectrum) for the Galerkin and the full GPE. This broad peak clearly represents a continuous spectrum, typical to non-integrable systems, that give rise to dynamical chaos and ergodicity \cite{Moon}.

%For the model with the harmonic-oscillator potential, the Galerkin approximation- and GPE-produced results are also virtually identical. These results are drastically different from those for the box-potential model, demonstrating a spectrum consisting of extremely sharp peaks, which may be definitely categorized as a \emph{%practically discrete} spectrum, thus representing \emph{quasi-integrable dynamics}. The peaks evident in the spectrum can be immediately identified as located at the above-mentioned frequencies $\omega _{d}$ and $\omega _{s}$ of the dipole mode of the excitations of the condensate as a whole [see Eq.~(% \ref{sx})], and shuttle oscillations of the DS [see Eq.~(\ref{ds})].

The finite width of the peaks representing the harmonic-oscillator potential
is attributed to numerical accuracy, the inherent frequency resolution of
the discrete Fourier transform for total simulation time $T$ being $\Delta
\omega =2\pi /T $. In the present case, $\Delta \omega \approx 0.0005$, and,
indeed, the width of the peaks is equal to $2\Delta \omega $.
%\large{[The main result of the paper is illustrated by the single figure, Fig.~\ref{fig:integ}. It will be very %helpful to include at least one more illustration. A natural one may correspond to a symmetric motion of two DSs, %periodically colliding at the center. I guess you have some plots like that, which may be used to additionally %uphold the main claims. \tom{issue is that two dark solitons can't be simulated in the harmonic oscillator case %without going to many more modes...}]}

%We have also performed the analysis for higher amplitudes of the Galerkin
%expansion (up to the $M=16$ considered in this work), finding perfect
%agreement between the Galerkin approximation and full GPE results, with
%discrete and continuous spectra obtained in the models with the
%harmonic-oscillator and box potentials, respectively.

\begin{figure}[tbp]
\centering
\hspace{-0.2cm} Harmonic trap \hspace{5cm} Box trap %
\includegraphics[scale=1.1]{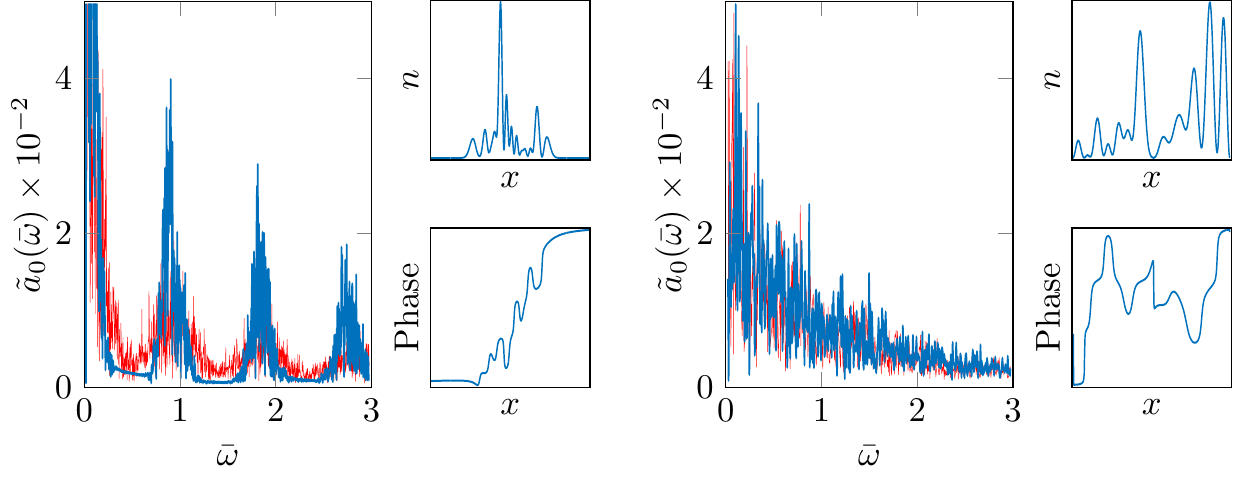}\newline
\caption{The first amplitude of the Galerkin expansion, with a set of
random-field amplitudes generated for the initial condition in both models,
with the harmonic-oscillator (left) and box (right) trapping potentials. The
power spectra are produced by the projection of the GPE simulations (blue)
onto the first eigenmode of the Galerkin basis. Simulations of the the
Galerkin system produced similar results, which are consistent with the
findings produced by the full GPE simulations(red). Other parameters are the
same as in Fig.~\protect\ref{fig:integ}.}
\label{fig:noise}
\end{figure}

A physically relevant situation, when a random initial condition is
generated for each coefficient $j=1,\dots,M$ such that $a_j(0)=x_j\exp%
\left(iy_j\right)$, where $x_j$ is an observation from the random variable $%
X\sim U(-1,1)$ and $y_j$ from $Y\sim U(0,2\pi)$ (taking care to renormalise
according to Eq.~(\ref{eqn:N})), has been tested too. Typical examples of
power spectra found in this case are displayed, for both the
harmonic-oscillator and box traps, in Fig.~\ref{fig:noise}. It is seen that
both models keep the character of their dynamics, corresponding to the
quasi-discrete and continuous spectra, respectively, in the presence of the
strong random component in the input. Thus, the quasi-integrability of the
GPE with the harmonic-oscillator potential is a robust property. Figure \ref%
{fig:noise} also shows the density and phase profiles of the initial
condition for both wave-fields. This initial condition is akin to a
soliton-gas configuration, generated by summation of several dark soliton
solutions with random position, phase and velocity. Using this comparison
we can describe the nature of the arising peaks. Considering the dynamics
in the harmonic-oscillator case, shallow (fast) solitons have an oscillation
frequency close to $\omega_s\approx\omega_x$, whereas deep (slow) solitons
near the condensate centre have an oscillation frequency near to $%
\omega_s\approx\omega_x/\sqrt{2}$, as shown previously. In the dynamics
ensuing from this highly nonequilibrium initial state the power spectra
displays a complicated mixing of these modes, with an envelope of spectra
centred around their average, $\omega\approx0.85\,\omega_x$. Similar to Fig.~%
\ref{fig:integ} the envelopes at larger $\omega$ are due to mixing of these
frequencies.

%, the results are precisely the same, as concerns the fact that the Galerkin approximation predictions exactly match the GPE simulations.  %Thus our main and general conclusion here is that the box-potential %model generates continuous spectra with a broad peak, while the harmonic-oscillator always %gives rise to quasi-discrete spectra with sharp peaks, located precisely at t%he same positions as in the left panel of Fig.~\ref{fig:integ}. While only one generic example of the spectra is explicitly shown here, all other cases considered give rise to exactly the same conclusions: first, the Galerkin approximation and the full GPE simulations produce virtually identical results for both models, with the harmonic-oscillator and box potentials. Second, and most important, the power spectra are built of very sharp quasi-discrete peaks in the harmonic-oscillator model, and, on the contrary, feature broad continuous spectra in the model with the box potential.
%{\LARGE [If the above suggestion is implemented, we will have not one generic example", as mentioned here, but two.]}

\section{Discussion}

Previous numerical simulations based on the 1D GPE have revealed shuttle
oscillations of dark solitons in the harmonic-oscillator potential. In the
course of the periodic motion, the dark soliton reversibly emits
small-amplitude waves (\textquotedblleft sound"), being able to fully
reabsorb them. No chaotization was observed in the course of indefinitely
long simulations of this model. On the contrary to that, GPE simulations
with other types of trapping potentials exhibit irreversible evolution and
an eventual trend to the onset of dynamical chaos (wave-function
\textquotedblleft turbulence") \cite{parker_2010,sciacca}. To explain this
phenomenology, we have first derived a finite-mode dynamical system, in the
form of the Galerkin approximation, based on the truncated expansion of the
wave function, governed by the GPE (Gross-Pitaevskii equation) with the
repulsive cubic nonlinearity, over the set of eigenmodes of the
corresponding linear Schr\"{o}dinger equation. The comparison of results
produced by the Galerkin approximation to those of full GPE simulations
shows that the Galerkin approximation for the model with the
harmonic-oscillator potential, with $M=16$ modes, reproduces the full
solutions virtually exactly (with fidelity indistinguishable from $1$, see Appendix B) for
indefinitely long evolution times. In the case of the box potential, the
Galerkin approximation with $M=16$ also provides a high accuracy, although,
eventually, there emerges a deviation from the GPE solutions at large
evolution times. The main finding is that generic trajectories of the
Galerkin approximation derived for the model with the harmonic-oscillator
potential produce a discrete power spectrum (up to the accuracy of the
numerically implemented Fourier transform), which is a remarkable
manifestation of the conjectured quasi-integrability. This finding (which
remains true in the presence of a strong random-noise component in the
input) strongly suggests that, in the underlying dynamical system, virtually
all trajectories wind upon invariant tori, only an extremely small share of
the tori (if any) being destroyed. It remains a challenge to understand the
quasi-integrability of the GPE with the harmonic-oscillator potential at a
deeper mathematical level than the explanation offered by the present
analysis.

On the other hand, both direct simulations of the GPE and the Galerkin
approximation for the model with the box potential produce a continuous
power spectrum, in the form of a very broad peak, which clearly implies that
the latter system is subject to the (rather slow) onset of chaotization.
Thus, this work puts forward an open question concerning the nature of the
non-integrable dynamics in the truncated version of the formally integrable
system~\cite{Maxim}.

It may also be interesting to perform a similar analysis to the one
performed here for the case of \emph{attractive} nonlinearity ($\sigma =-1$
in Eq.~(\ref{L})) focussing on shuttle oscillations of a trapped bright
soliton (see also the related work in Ref.~\cite{martin_PRA}), as well as
for recurrent collisions between two (or several) solitons (the latter
setting was experimentally realized in the self-attracting BEC\ \cite{Randy}%
). Furthermore, for both cases of the self-repulsion and attraction, the
analysis may be extended to a two-component GPE with equal strengths of the
self- and cross-interactions, which, in the free space, corresponds to the
integrable Manakov's system \cite{Manakov}. This system remains integrable
too if it includes the Rabi coupling, i.e., linear interconversion between
the components \cite{Tratnik}, which is thus also an appropriate subject for
the consideration. The Manakov's system finds the well-known realization in
terms of the two-component BEC mixtures \cite{BEC}. Moreover, this
methodology may enable insight into the stability and dynamics of dark
solitons within the nonlocal dipolar GPE, an equation which can be realised
experimentally through BECs of atoms which possess strong magnetic dipoles;
while the nonlocality breaks the integrability of this governing mean-field
equation, dark solitons were also found to show quasi-integrable dynamics,
both in homogeneous \cite{Pawlowski,Bland1,Edmonds} and trapped systems \cite%
{Bland2}.

Interesting questions are also expected to arise in the development of the
Galerkin approximation for the two-dimensional (2D) GPE with an isotropic
harmonic-oscillator potential (see also Ref.~\cite{Chinese} for
multidimensional Schr\"{o}dinger equations with generalized nonlinearities
and damping). In particular, it is known that the 2D model with the
attractive nonlinearity and harmonic-oscillator trapping potential makes the
trapped fundamental solitons completely stable (against the critical
collapse in the 2D space \cite{collapse}), and provides for partial
stabilization of vortex solitons with topological charge $1$ against the
collapse and splitting \cite{Dum}.
%In the 2D setting, the full set of the modes for the Galerkin expansion is
%provided by the commonly known harmonic oscillator eigenstates with integer values of the
%angular momentum, $l=0,1,2,...$ .
The investigation of the 2D model may be interesting also for the reason
that the 2D GPE in the free space is not integrable, the question being if
the harmonic-oscillator confinement may induce a quasi-integrability in this
case.

%In this work, the scheme for the calculation of the Galerkin approximation
%coefficients was implemented by projecting numerical solutions of the GPE\
%onto the respective modes, see Eq.~(\ref{fid}). While the excellent
%agreement shown in our work for the 1D setting allows little room for
%improvement, the projection scheme is likely to prove beneficial in higher
%dimensions, or in mode complex systems (in particular, multi-component ones).

Data supporting this work is openly available under an `Open Data Commons
Open Database License' \cite{data}.

\section*{Acknowledgments}

We acknowledge valuable discussions with Tom Billam, Maxim Olshanii, and
Alexander Its. N.G.P. acknowledges funding from the Engineering and Physical
Sciences Research Council (Grant No. EP/M005127/1). T. B. acknowledges
support from Engineering and Physical Sciences Research Council. The work of
B.A.M. on this project was carried out in the framework of the visiting
professorship provided by the Newcastle University. This author also
acknowledges support provided by grant No.~2015616 from the joint program in
physics between the NSF and Binational (US-Israel) Science Foundation, and
by grant No.~1286/17 from the Israel Science Foundation.

\appendix

\section{The four-mode truncated Hamiltonian and dynamical equations}

\setcounter{section}{1}

In this appendix we provide an explicit example of the dynamical system
produced by the Galerkin approximation with $M=4$ modes in the model with
the harmonic-oscillator potential. The Hermite polynomials required to
construct the corresponding Galerkin approximation ansatz are
\[
H_{0}(x)=1,~H_{1}(x)=2x,~H_{2}(x)=2(2x^{2}-1),~H_{3}(x)=4x\left(
2x^{2}-3\right) .
\]%
Calculation of the quartic term in the corresponding Lagrangian (\ref{L})
leads to Hamiltonian (\ref{eq:H}) in the following form: {%
\setlength{\mathindent}{0pt}
\begin{eqnarray*}
H=\frac{\sigma }{2\sqrt{2\pi }} &\Bigg[&\left\vert a_{0}\right\vert ^{4}+%
\frac{3}{4}\left\vert a_{1}\right\vert ^{4}+\frac{41}{64}\left\vert
a_{2}\right\vert ^{4}+\frac{147}{256}\left\vert a_{3}\right\vert
^{4}+2|a_{0}|^{2}|a_{1}|^{2}+\frac{3}{2}|a_{0}|^{2}|a_{2}|^{2}+\frac{7}{4}%
|a_{1}|^{2}|a_{2}|^{2} \\
&~&+\frac{5}{4}|a_{0}|^{2}|a_{3}|^{2}+\frac{11}{8}|a_{1}|^{2}|a_{3}|^{2}+%
\frac{51}{32}|a_{2}|^{2}|a_{3}|^{2}+\frac{\sqrt{3}}{4}\left(
a_{1}a_{2}a_{0}^{\ast }a_{3}^{\ast }+a_{0}a_{1}^{\ast }a_{2}^{\ast
}a_{3}\right) \\
&~&+\frac{5\sqrt{3}}{16\sqrt{2}}\left( a_{2}^{2}a_{1}^{\ast }a_{3}^{\ast
}+a_{1}\left( a_{2}^{\ast }\right) ^{2}a_{3}\right) +\frac{1}{2\sqrt{2}}%
\left( a_{1}^{2}a_{0}^{\ast }a_{2}^{\ast }+a_{0}a_{2}\left( a_{1}^{\ast
}\right) ^{2}\right) \\
&~&+\frac{5}{16}\left( a_{0}^{2}\left( a_{3}^{\ast }\right) ^{2}+\left(
a_{0}^{\ast }\right) ^{2}a_{3}^{2}\right) +\frac{11}{32}\left(
e^{4it}a_{1}^{2}\left( a_{3}^{\ast }\right) ^{2}+e^{-4it}\left( a_{1}^{\ast
}\right) ^{2}a_{3}^{2}\right) \\
&~&+\frac{51}{128}\left( e^{2it}a_{2}^{2}\left( a_{3}^{\ast }\right)
^{2}+e^{-2it}\left( a_{2}^{\ast }\right) ^{2}a_{3}^{2}\right) +\frac{1}{16%
\sqrt{2}}\left( e^{4it}a_{0}a_{2}\left( a_{3}^{\ast }\right)
^{2}+e^{-4it}a_{0}^{\ast }a_{2}^{\ast }a_{3}^{2}\right) \\
&~&+\frac{3\sqrt{3}}{32\sqrt{2}}\left( e^{2it}a_{1}a_{3}^{\ast
}|a_{3}|^{2}+e^{-2it}a_{1}^{\ast }|a_{3}|^{2}a_{3}\right) +\frac{1}{2}\left(
e^{2it}a_{0}^{2}\left( a_{1}^{\ast }\right) ^{2}+e^{-2it}a_{1}^{2}\left(
a_{0}^{\ast }\right) ^{2}\right) \\
&~&+\frac{1}{\sqrt{2}}\left( e^{2it}a_{0}|a_{1}|^{2}a_{2}^{\ast
}+e^{-2it}|a_{1}|^{2}a_{2}a_{0}^{\ast }\right) -\frac{1}{\sqrt{2}}\left(
e^{2it}|a_{0}|^{2}a_{0}a_{2}^{\ast }+e^{-2it}|a_{0}|^{2}a_{2}a_{0}^{\ast
}\right) \\
&~&-\sqrt{\frac{3}{2}}\left( e^{2it}a_{1}|a_{0}|^{2}a_{3}^{\ast
}+e^{-2it}|a_{0}|^{2}a_{1}^{\ast }a_{3}\right) +\frac{7}{16}\left(
e^{2it}a_{1}^{2}\left( a_{2}^{\ast }\right) ^{2}+e^{-2it}a_{2}^{2}\left(
a_{1}^{\ast }\right) ^{2}\right) \\
&~&+\frac{1}{8\sqrt{2}}\left( e^{2it}a_{0}|a_{2}|^{2}a_{2}^{\ast
}+e^{-2it}|a_{2}|^{2}a_{2}a_{0}^{\ast }\right) -\frac{\sqrt{3}}{2\sqrt{2}}%
\left( e^{4it}a_{0}^{2}a_{1}^{\ast }a_{3}^{\ast }+e^{-4it}a_{1}\left(
a_{0}^{\ast }\right) ^{2}a_{3}\right) \\
&~&+\frac{5\sqrt{3}}{8\sqrt{2}}\left( e^{2it}a_{1}|a_{2}|^{2}a_{3}^{\ast
}+e^{-2it}|a_{2}|^{2}a_{1}^{\ast }a_{3}\right) -\frac{\sqrt{3}}{4\sqrt{2}}%
\left( e^{2it}|a_{1}|^{2}a_{1}a_{3}^{\ast }+e^{-2it}|a_{1}|^{2}a_{1}^{\ast
}a_{3}\right) \\
&~&+\frac{\sqrt{3}}{4}\left( e^{4it}a_{0}a_{1}a_{2}^{\ast }a_{3}^{\ast
}+e^{-4it}a_{2}a_{0}^{\ast }a_{1}^{\ast }a_{3}\right) +\frac{3}{8}\left(
e^{4it}a_{0}^{2}\left( a_{2}^{\ast }\right) ^{2}+e^{-4it}a_{2}^{2}\left(
a_{0}^{\ast }\right) ^{2}\right) \\
&~&+\frac{\sqrt{3}}{4\sqrt{2}}\left( e^{2it}a_{0}a_{2}a_{1}^{\ast
}a_{3}^{\ast }+e^{-2it}a_{1}a_{0}^{\ast }a_{2}^{\ast }a_{3}\right) +\frac{1}{%
8\sqrt{2}}\left( e^{2it}a_{0}a_{2}^{\ast
}|a_{3}|^{2}+e^{-2it}a_{2}a_{0}^{\ast }|a_{3}|^{2}\right) \Bigg].
\end{eqnarray*}%
}%
\[
\]%
Finally, substituting this Hamiltonian in Euler-Lagrange equations (\ref{EL}%
), we arrive at the following dynamical system with four degrees of freedom:
{\setlength{\mathindent}{0pt}
\begin{eqnarray*}
i\frac{2\sqrt{2\pi }}{\sigma }\frac{da_{0}}{dt} &=&2a_{0}^{\ast }a_{0}^{2}-%
\frac{1}{\sqrt{2}}a_{2}^{\ast }e^{2it}a_{0}^{2}-\sqrt{2}a_{2}a_{0}^{\ast
}e^{-2it}a_{0}-\sqrt{\frac{3}{2}}a_{3}a_{1}^{\ast }e^{-2it}a_{0}-\sqrt{\frac{%
3}{2}}a_{1}a_{3}^{\ast }e^{2it}a_{0} \\
&~&+2a_{1}a_{1}^{\ast }a_{0}+\frac{3}{2}a_{2}a_{2}^{\ast }a_{0}+\frac{5}{4}%
a_{3}a_{3}^{\ast }a_{0}+\frac{5}{8}a_{3}^{2}a_{0}^{\ast }e^{-6it}+\frac{3}{4}%
a_{2}^{2}a_{0}^{\ast }e^{-4it}-\sqrt{\frac{3}{2}}a_{1}a_{3}a_{0}^{\ast
}e^{-4it} \\
&~&+\frac{1}{4}\sqrt{3}a_{2}a_{3}a_{1}^{\ast }e^{-4it}+\frac{1}{16\sqrt{2}}%
a_{3}^{2}a_{2}^{\ast }e^{-4it}+a_{1}^{2}a_{0}^{\ast }e^{-2it}+\frac{1}{\sqrt{%
2}}a_{1}a_{2}a_{1}^{\ast }e^{-2it} \\
&~&+\frac{1}{8\sqrt{2}}a_{2}^{2}a_{2}^{\ast }e^{-2it}+\frac{1}{4}\sqrt{3}%
a_{1}a_{3}a_{2}^{\ast }e^{-2it}+\frac{1}{8\sqrt{2}}a_{2}a_{3}a_{3}^{\ast
}e^{-2it}+\frac{1}{2\sqrt{2}}a_{1}^{2}a_{2}^{\ast }+\frac{1}{4}\sqrt{3}%
a_{1}a_{2}a_{3}^{\ast }, \\
i\frac{2\sqrt{2\pi }}{\sigma }\frac{da_{1}}{dt} &=&a_{1}^{\ast
}e^{2it}a_{0}^{2}-\frac{1}{2}\sqrt{\frac{3}{2}}a_{3}^{\ast }e^{4it}a_{0}^{2}-%
\sqrt{\frac{3}{2}}a_{3}a_{0}^{\ast }e^{-2it}a_{0}+\frac{1}{\sqrt{2}}%
a_{1}a_{2}^{\ast }e^{2it}a_{0} \\
&~&+\frac{1}{4}\sqrt{3}a_{2}a_{3}^{\ast }e^{2it}a_{0}+2a_{1}a_{0}^{\ast
}a_{0}+\frac{1}{\sqrt{2}}a_{2}a_{1}^{\ast }a_{0}+\frac{1}{4}\sqrt{3}%
a_{3}a_{2}^{\ast }a_{0}+\frac{1}{4}\sqrt{3}a_{2}a_{3}a_{0}^{\ast }e^{-4it} \\
&~&+\frac{11}{16}a_{3}^{2}a_{1}^{\ast }e^{-4it}+\frac{1}{\sqrt{2}}%
a_{1}a_{2}a_{0}^{\ast }e^{-2it}+\frac{7}{8}a_{2}^{2}a_{1}^{\ast }e^{-2it}-%
\frac{1}{2}\sqrt{\frac{3}{2}}a_{1}a_{3}a_{1}^{\ast }e^{-2it} \\
&~&+\frac{5}{8}\sqrt{\frac{3}{2}}a_{2}a_{3}a_{2}^{\ast }e^{-2it}+\frac{3}{32}%
\sqrt{\frac{3}{2}}a_{3}^{2}a_{3}^{\ast }e^{-2it}-\frac{1}{4}\sqrt{\frac{3}{2}%
}a_{1}^{2}a_{3}^{\ast }e^{2it}+\frac{3}{2}a_{1}^{2}a_{1}^{\ast } \\
&~&+\frac{7}{4}a_{1}a_{2}a_{2}^{\ast }+\frac{5}{16}\sqrt{\frac{3}{2}}%
a_{2}^{2}a_{3}^{\ast }+\frac{11}{8}a_{1}a_{3}a_{3}^{\ast }, \\
i\frac{2\sqrt{2\pi }}{\sigma }\frac{da_{2}}{dt} &=&\frac{3}{4}a_{2}^{\ast
}e^{4it}a_{0}^{2}-\frac{1}{\sqrt{2}}a_{0}^{\ast }e^{2it}a_{0}^{2}+\frac{1}{%
\sqrt{2}}a_{1}a_{1}^{\ast }e^{2it}a_{0}+\frac{1}{4\sqrt{2}}a_{2}a_{2}^{\ast
}e^{2it}a_{0} \\
&~&+\frac{1}{8\sqrt{2}}a_{3}a_{3}^{\ast }e^{2it}a_{0}+\frac{1}{4}\sqrt{3}%
a_{1}a_{3}^{\ast }e^{4it}a_{0}+\frac{3}{2}a_{2}a_{0}^{\ast }a_{0}+\frac{1}{4}%
\sqrt{3}a_{3}a_{1}^{\ast }a_{0}+\frac{1}{16\sqrt{2}}a_{3}^{2}a_{0}^{\ast
}e^{-4it} \\
&~&+\frac{1}{8\sqrt{2}}a_{2}^{2}a_{0}^{\ast }e^{-2it}+\frac{1}{4}\sqrt{3}%
a_{1}a_{3}a_{0}^{\ast }e^{-2it}+\frac{5}{8}\sqrt{\frac{3}{2}}%
a_{2}a_{3}a_{1}^{\ast }e^{-2it}+\frac{51}{64}a_{3}^{2}a_{2}^{\ast }e^{-2it}
\\
&~&+\frac{7}{8}a_{1}^{2}a_{2}^{\ast }e^{2it}+\frac{5}{8}\sqrt{\frac{3}{2}}%
a_{1}a_{2}a_{3}^{\ast }e^{2it}+\frac{1}{2\sqrt{2}}a_{1}^{2}a_{0}^{\ast }+%
\frac{7}{4}a_{1}a_{2}a_{1}^{\ast }+\frac{41}{32}a_{2}^{2}a_{2}^{\ast } \\
&~&+\frac{5}{8}\sqrt{\frac{3}{2}}a_{1}a_{3}a_{2}^{\ast }+\frac{51}{32}%
a_{2}a_{3}a_{3}^{\ast }, \\
i\frac{2\sqrt{2\pi }}{\sigma }\frac{da_{3}}{dt} &=&-\frac{1}{2}\sqrt{\frac{3%
}{2}}a_{1}^{\ast }e^{4it}a_{0}^{2}+\frac{5}{8}a_{3}^{\ast }e^{6it}a_{0}^{2}-%
\sqrt{\frac{3}{2}}a_{1}a_{0}^{\ast }e^{2it}a_{0}+\frac{1}{4}\sqrt{3}%
a_{2}a_{1}^{\ast }e^{2it}a_{0} \\
&~&+\frac{1}{8\sqrt{2}}a_{3}a_{2}^{\ast }e^{2it}a_{0}+\frac{1}{4}\sqrt{3}%
a_{1}a_{2}^{\ast }e^{4it}a_{0}+\frac{1}{8\sqrt{2}}a_{2}a_{3}^{\ast
}e^{4it}a_{0}+\frac{5}{4}a_{3}a_{0}^{\ast }a_{0}+\frac{1}{8\sqrt{2}}%
a_{2}a_{3}a_{0}^{\ast }e^{-2it} \\
&~&+\frac{3}{32}\sqrt{\frac{3}{2}}a_{3}^{2}a_{1}^{\ast }e^{-2it}-\frac{1}{4}%
\sqrt{\frac{3}{2}}a_{1}^{2}a_{1}^{\ast }e^{2it}+\frac{5}{8}\sqrt{\frac{3}{2}}%
a_{1}a_{2}a_{2}^{\ast }e^{2it}+\frac{51}{64}a_{2}^{2}a_{3}^{\ast }e^{2it} \\
&~&+\frac{3}{16}\sqrt{\frac{3}{2}}a_{1}a_{3}a_{3}^{\ast }e^{2it}+\frac{11}{16%
}a_{1}^{2}a_{3}^{\ast }e^{4it}+\frac{1}{4}\sqrt{3}a_{1}a_{2}a_{0}^{\ast }+%
\frac{5}{16}\sqrt{\frac{3}{2}}a_{2}^{2}a_{1}^{\ast } \\
&~&+\frac{11}{8}a_{1}a_{3}a_{1}^{\ast }+\frac{51}{32}a_{2}a_{3}a_{2}^{\ast }+%
\frac{147}{128}a_{3}^{2}a_{3}^{\ast }.
\end{eqnarray*}%
} The MATLAB code used to solve the dynamical system for $M=4, 8, 16$ in the
harmonic-oscillator and box potentials is available online \cite{ data}.
\newline

\section{Exploring the accuracy of the Galerkin approximation}

Here we characterise the ability of the Galerkin approximation to capture
the stationary ground state of the system, as a function of the mode number $%
M$ and norm $N$. Using a stationary solution $\psi _{\mathrm{GPE}}$ of the
GPE (obtained by means of the imaginary-time propagation \cite{primer}),
Galerkin approximation amplitudes $a_{j}$ (see Eqs.~(\ref{ga_harm}) and (\ref%
{eqn:ga_box})) are calculated by
projecting $\psi _{\mathrm{GPE}}$ onto the set of eigenstates of the
respective linearized Schr\"{o}dinger equation,
\begin{equation}
a_{j}(0)=\int_{\mathcal{V}}\mathrm{d}x~\psi _{j}(x)\psi _{\mathrm{GPE}}(x),
\label{fid}
\end{equation}%
where $\mathcal{V}$ is the actual range of $x$ for the harmonic-oscillator
potential, and interval $0<x<L$ for the box.
%This integral defines a measure of the overlap of the two states, hereby referred to as a the fidelity of the jth state.
Using the amplitudes given by Eq.~(\ref{fid}) to construct the Galerkin
approximation wave function $\psi _{\mathrm{GA}}(x,0)$, as per Eqs. (\ref%
{ga_harm}) and (\ref{eqn:ga_box}), we define the \textit{fidelity} $F$ of
the approximation as
\begin{equation}
F=\frac{1}{N}\int_{\mathcal{V}}\mathrm{d}x~\psi _{\mathrm{GA}}(x,0)\psi _{%
\mathrm{GPE}}(x),  \label{F}
\end{equation}%
where $F=1$ ($F=0$) corresponds to two identical (mutually orthogonal) wave
functions.
\begin{figure}[h]
\centering
\hspace{0.9cm} Harmonic trap \hspace{4.5cm} Box trap
\par
\includegraphics[scale=1.1]{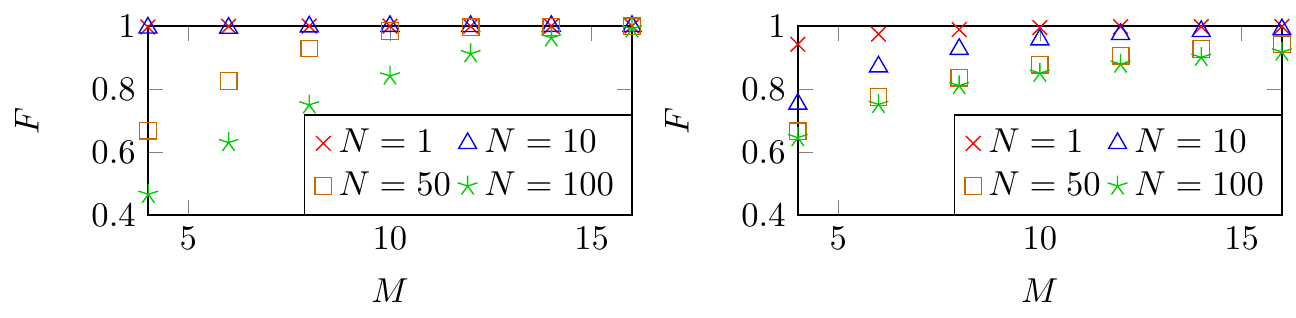} \newline
\hspace{-0.1cm}
\includegraphics[scale=1.1]{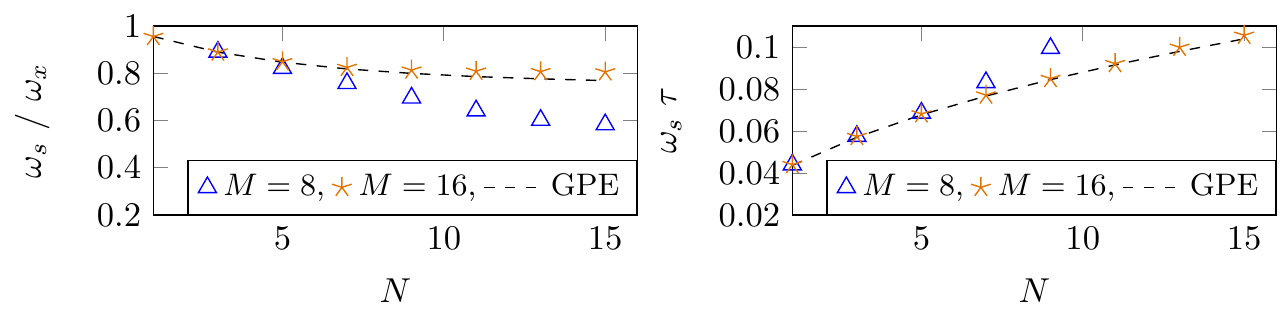}
\caption{(Top row) Fidelity of the ground-state wave function vs.~the number
of modes, $M$, kept in the Galerkin approximation, shown for the
harmonic-oscillator (left) and box (right) traps, with different values of
norm $N$.
(Bottom row) Effect of the increasing number of modes in the Galerkin
approximation on its accuracy, estimated by the comparison of the frequency
of the shuttle oscillations of the dark soliton with results of the GPE
simulations. Other parameters used here are the same as in Fig.~\protect\ref%
{fig:osc_eg}.}
\label{fig:fid}
\end{figure}

Figure \ref{fig:fid} (top row) shows how the number of modes affects the
fidelity of the initial Galerkin approximation wave function, for different
norms $N$. %Each point was found
%by solving the GPE in imaginary time, and then substituting the so-obtained $%
%\psi _{\mathrm{GPE}}$ into Eq.~(\ref{F}), along with the $M$-mode truncated
%ansatz, given by Eq.~(\ref{ga_harm}) or (\ref{eqn:ga_box}).
In the case of the harmonic-oscillator trap and $M=16$, the fidelity is
virtually exactly $F=1 $, implying an almost perfect GPE-Galerkin
approximation overlap for all norms considered. For the box trap, the
fidelity is still good but poorer than for the harmonic-oscillator; even in
the case of $N=1$ (weak nonlinearity), $\psi _{\mathrm{GPE}}$ for the box is
not perfectly approximated by the truncation with $M<10$.
%In what followsbelow, we focus on three cases: $M=4,8$ and $16$, for both the
%harmonic-oscillator and box traps.

Finally we explore the validity of the Galerkin model for recreating
dynamical simulations of the GPE. Figure \ref{fig:fid} (bottom row) explores
the role of the mode truncation in the Galerkin approximation, by direct
comparison of the oscillation frequencies, extracted from the GPE
simulations, and their counterparts, predicted by the Galerkin
approximation, as a function of the total norm. The results clearly show
that the increasing number of modes, $M$, improves the Galerkin
approximation\ accuracy for all norms considered. \newline

%\bibitem{fletche_1984} Fletche C A J 1984 \textit{Computational Galerkin
%Methods} (Springer: Berlin)

\end{document}